\pdfoutput=1
\PassOptionsToPackage{dvipsnames,table}{xcolor}
\documentclass[11pt]{article}

\usepackage[final]{acl}

\usepackage{times}
\usepackage{latexsym}

\usepackage[T1]{fontenc}

\usepackage[utf8]{inputenc}

\usepackage{microtype}
\usepackage[dvipsnames,table]{xcolor}

\usepackage{graphicx}
\newcommand{\quotes}[1]{``#1''}

\usepackage{graphicx}
\usepackage{amsmath}
\usepackage{booktabs}
\usepackage{algorithm}
\usepackage{comment}
\usepackage{bbold}
\usepackage{booktabs, multirow}
\usepackage{graphicx,array}
\usepackage{tabularx}

\usepackage{color,soul,xspace}
\usepackage{hyperref}
\usepackage{breakurl}
\usepackage{xurl}
\usepackage{comment}
\usepackage{epsfig}
\usepackage{subfig}
\usepackage{mathrsfs,mdwlist,enumitem}

\title{A Survey on Patent Analysis: From NLP to Multimodal AI}

\author{First Author \\
  Affiliation / Address line 1 \\
  Affiliation / Address line 2 \\
  Affiliation / Address line 3 \\
  \texttt{email@domain} \\\And
  Second Author \\
  Affiliation / Address line 1 \\
  Affiliation / Address line 2 \\
  Affiliation / Address line 3 \\
  \texttt{email@domain} \\}
\author{%
  Homaira Huda Shomee \qquad Zhu Wang  \qquad Sathya N. Ravi \qquad Sourav Medya\\
  Department of Computer Science, University of Illinois Chicago\\
  \texttt{\{hshome2,zwang260,sathya,medya\}@uic.edu} \\
  }

\begin{document}
\maketitle


\begin{abstract}
Recent advances in Pretrained Language Models (PLMs) and Large Language Models (LLMs) have demonstrated transformative capabilities across diverse domains. The field of patent analysis and innovation is not an exception, where natural language processing (NLP) techniques presents opportunities to streamline and enhance important tasks---such as patent classification and patent retrieval---in the patent cycle. This not only accelerates the efficiency of patent researchers and applicants, but also opens new avenues for technological innovation and discovery. Our survey provides a comprehensive summary of recent NLP-based methods---including multimodal ones---in patent analysis. We also introduce a novel taxonomy for categorization based on tasks in the patent life cycle, as well as the specifics of the methods. This interdisciplinary survey aims to serve as a comprehensive resource for researchers and practitioners who work at the intersection of NLP, Multimodal AI, and patent analysis, as well as patent offices to build efficient patent systems. 
\end{abstract}

\section{Introduction}


The growing complexity and volume of textual data across various domains have driven significant advancements in NLP, particularly through PLMs \cite{BERT} and LLMs \cite{gpt2}. The field of patents and technological innovation is not an exception. This advancement can streamline complex patent-related tasks such as classification, retrieval, and valuation prediction. For instance, for patent examination, patent offices often rely only on the examiner to judge whether a technology is innovative enough and, thus, patentable. However, it is challenging for the human examiner to stay updated on various domains due to the exponential growth in technology and apply the knowledge during evaluation. This intersection of NLP, Multimodal AI, and patent processes can accelerate the efficiency of the patent systems---patent reviewers as well as applicants---and help in a faster technological innovation to benefit our society.\\
The patent application and granting process involves complex textual analysis tasks that require significant human effort for both applicants and reviewers. To streamline this, NLP techniques can be helpful, particularly in patent classification, retrieval, and quality analysis \cite{krestel2021survey}. Patent classification can benefit from multi-label classification tools for the hierarchical schemes: International Patent Classification (IPC) and the Cooperative Patent Classification \cite{haghighian2022patentnet,althammer2021linguistically}. To evaluate novelty and avoid infringement, the patent retrieval task becomes important while filing or reviewing a new patent application. On the other hand, quality analysis also requires a substantial amount of effort. NLP-based representation learning methods can be useful in both tasks \cite{chung2020early,lin2018patent}. Lastly, recent advanced LLMs can generate accurate and technical language descriptions for patents and, thus, are useful to optimize human resources and precision in patent writing \cite{lee2020patent}. \\
The existing patent surveys in the literature \cite{gomez2014survey, ali2024innovating,krestel2021survey,hanbury2011patent,casola2022summarization} do not cover the recent studies in this area and fail to show the trends and methods in task specific manner. We introduce a novel taxonomy to categorize the methods based on the relevant tasks and the nature of the methods. Our taxonomy provides an in-depth view of the methods being used in specific tasks. Moreover, it captures the recent trends of using advanced methods (e.g., LLMs) that are missing from the existing surveys. This will be beneficial for researchers who aim to build task-specific methods. 

\begin{figure*}[ht]
\centering
\vspace{-3mm}
\includegraphics[width=0.95\textwidth]{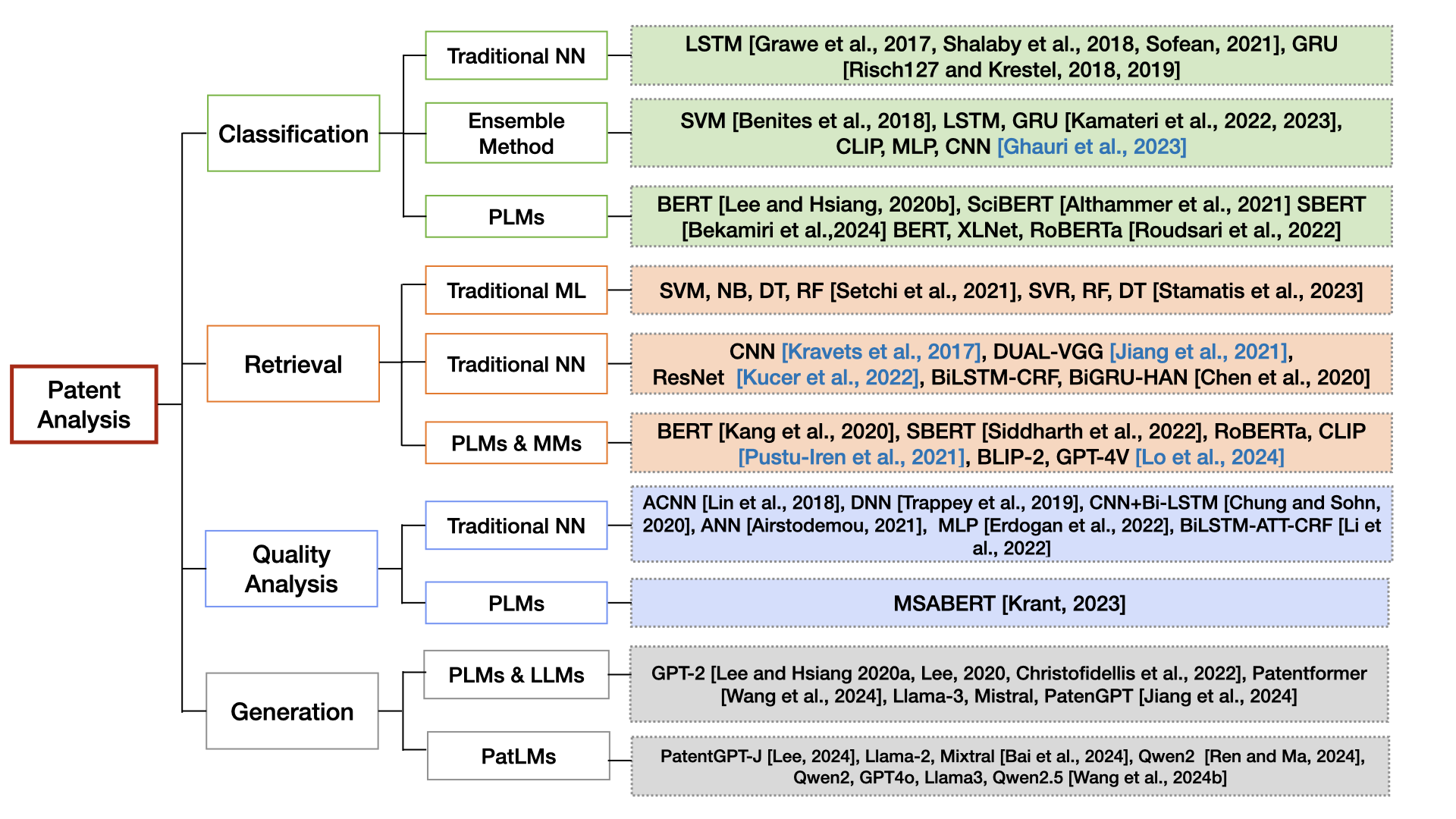}
\vspace{-1mm}
\caption {\label{fig:schema}%
The schema of the main organization with the methods in each patent-related task. We summarize the methods for four individual tasks: patent classification, retrieval, quality analysis, and generation. \quotes{NN}, \quotes{MMs}, \quotes{PLMs}, and \quotes{PatLMs} denote neural networks, multimodal models, pre-trained language models, and patent language models, respectively. The works that use patent images are written in blue.}
\vspace{-2mm}
\end{figure*}

\textbf{Overview.} 
Fig. \ref{fig:schema} provides the hierarchical organization of patent tasks and methods.  We
organize the survey as follows: Sec. \ref{sec:background} provides background, Sec. \ref{sec:methods} summarizes the methods for individual tasks, and Sec. \ref{sec:future} provides future research directions. We maintain a GitHub repository for this survey at \href{https://github.com/hhshomee/AI4Patents-survey}{\texttt{AI4Patents-survey}}, which includes categorized papers and other relevant resources.


\section{Background}
\label{sec:background}

A patent grants the owner or holder exclusive rights to an invention and can be a novel product or a process that usually offers a unique method or technical solution. In exchange for this right, inventors must publicly disclose detailed information about their invention in a patent application. The United States Patent and Trademark Office (USPTO\footnote{\url{https://www.uspto.gov/}}) issues three types of patents: utility, design, and plant. In this work, we focus on utility and design patents, considering their importance in innovation across industries. Utility patents protect the rights related to how the invention works or is used. It provides the entitlement to the functionality of a product. On the other hand, design patents protect the right of the look of an invention and are intended to safeguard the form of a product. 
Here, we outline the relevant tasks.\\
 \textbf{Formulation. }We provide the problem formulations of these patent tasks in Appendix \ref{app:formulation}. \\
\textbf{Datasets. } We describe the common benchmark patent datasets in Appendix \ref{app:datasets}.

\subsection{Patent Classification}
Patent classification is an important but time-intensive task in the patent life cycle \cite{grawe2017automated,shalaby2018lstm,risch2018learning}. This involves a multi-label classification for patents where the classification scheme is hierarchical, and a patent can get multiple labels. There are two widely used patent classification systems: International Patent Classification (IPC) and the Cooperative Patent Classification (CPC). The IPC comprises 8 sections, 132 classes, 651 subclasses, 7590 groups, and 70788 subgroups in a hierarchical order (i.e., sections have classes and classes have subclasses, and so on). CPC is an expansion of IPC and is collaboratively administered by the European Patent Office (EPO) and the USPTO. It consists of around 250,000 classification entries and is divided into nine sections (A-H and Y), which are further broken down into classes, subclasses, groups, and subgroups\footnote{\url{https://www.cooperativepatentclassification.org/}}. Table \ref{tab:classification_example} (see Appendix) shows an example of CPC classification. 

\textbf{Challenges.} Patent classification is challenging due to its multi-class and multi-label nature. A single patent can be assigned multiple CPC/IPC codes, which makes the classification process complex. Additionally, the hierarchical structure of patent taxonomies introduces dependencies that require models to capture relationships between broad and fine-grained categories. Moreover, patent documents have various sections such as titles, abstracts, and claims---each contains different information. Given the extensive length of these full-text patent documents, identifying the most relevant sections for classification also poses a significant challenge.

\subsection{Patent Retrieval} 
Patent Retrieval (PR) \cite{Kravets2017/12,kang2020patent,chen2020deep,setchi2021artificial} focuses on developing methods to efficiently retrieve relevant patent documents and images based on specific search queries.  
PR plays a crucial role in identifying new patents related to new inventions. It is essential for evaluating novelty of a patent as well as ensuring that it does not infringe on existing patents. Moreover, patent image retrieval can serve as a source of inspiration for design.

\textbf{Challenges.} 
Patent retrieval tasks involve both text and image retrieval with unique challenges. Text retrieval is complex due to the use of similar words to describe new inventions; an invention can be described using various synonyms and phrasings which make it difficult to retrieve crucial information for patent infringement analysis. On the other hand, image retrieval is particularly challenging due to the nature of the images involved, which are typically black and white sketches, including numbers to describe the inventions. 
\subsection{Patent Quality analysis}
Businesses have shown great interest in evaluating patent value due to its significant impact in generating revenue and investment \cite{aristodemou2021identifying}. Investors usually aim to predict the future value of technological innovation from the target firm while making investment decisions. As a result, many companies hire professional patent analysts for quality analysis. This complex task demands substantial human effort as well as expertise in various domains \cite{lin2018patent}. The quality of a patent can be assessed using various measures, including the number of forward or backward citations, the number of claims, the grant lag, patent family size, the remaining lifetime of the
patent \cite{aristodemou2021identifying,erdogan2022predicting}.

\textbf{Challenges.} The challenge in analyzing patent quality is the ambiguity of the metrics to quantify the quality of a patent. Commonly used measures for the quality analysis are the number of citations (both forward and backward), the number of claims, and the grant lag. However, the weight of each of these measures remains unclear. Moreover, analyzing these information to perform a comprehensive study is non-trivial.

\subsection{Patent Generation}
Patents usually require a considerable amount of written text, which requires significant human resources. The patent generation task involves generating specific sections of a patent, such as abstract, independent claims, and dependent claims, based on instructions for an AI tool. Patent documents require precise and technical language to accurately describe the invention and its claims \cite{risch2021patentmatch}. AI-assisted patent generation will help automate the drafting process, which involves time, effort, and legal requirements. This will also reduce the amount of patent attorney time which will be a substantial cost saver.

\textbf{Challenges.} Though the patent document has certain structures, one major challenge is to evaluate the dependency---which can help in patent generation---among the parts of the patent. For instance, one part (e.g., abstract, claims) can be used as an input in a generative model (e.g., a LLM) to generate a different part of the patent.
 Additionally, it becomes non-trivial to construct effective instructions or prompts that guide the generation process. The generation also brings the question of evaluation of the generated content or text, i.e., how to judge whether the generated content is desired or not appropriate.

\section{Methods}
\label{sec:methods}
\vspace{-1mm}
We organize the important patent tasks that can benefit from recent advancements in NLP and Multimodal AI. An overview of important patent tasks is shown in Figure \ref{fig:overview} (Appendix \ref{app:overview}). The frequently used AI methods in the papers covered by this survey are in Table \ref{tab:summary_methods}. 
\subsection{Patent Classification}
In the literature, several models have been used to automate this process. We organize them based on the nature of the method into three major categories. Table \ref{tab:patent_classification} represents a summary of the methods for patent classification.
We present the evaluation metrics and the results in Table \ref{tab:patent_classification_results} in Appendix \ref{app:eval}. 
\begin{table*}[ht]
\caption{Studies on patent classification. Hierarchy levels for classification include Section, Class (white), Subclass (blue), Group, and Subgroup (grey). The color green represents the category of visualizations. Here, ADC denotes abstracts, descriptions, and claims, and TADC denotes titles, abstracts, descriptions, and claims. Table \ref{tab:patent_classification_results} provides more details on the performance in the Appendix.}  

\centering
\small
\resizebox{.9\textwidth}{!}{%

\begin{tabular}{cccc}\toprule
\textbf{Papers} & \textbf{Embeddings} & \textbf{Methods} &\textbf{Components}\\ 
\hline 
\cellcolor{gray!20}\cite{grawe2017automated} &  Word2Vec& Single layer LSTM & Description\\ 
\cellcolor{blue!20}\cite{shalaby2018lstm}&    Fixed Hierarchy Vectors    & LSTM & ADC\\
 \cellcolor{blue!20}\cite{risch2018learning}    &  FastText    & GRU& Full text\\
\cite{benites2018classifying}    & TF-IDF & SVM & Single Text Block
\\
\cellcolor{blue!20}\cite{risch2019domain}     & FastText&  GRU& Full text\\
\cellcolor{blue!20}\cite{2lee2020patent}& -- & BERT-base & Claim\\
\cellcolor{blue!20}\cite{althammer2021linguistically}& -- & BERT, SciBERT & Claim\\
\cellcolor{blue!20}\cite{sofean2021deep} & Word2Vec& Multiple LSTMs& Description\\
\cellcolor{blue!20}\cite{haghighian2022patentnet} & Word2Vec, FastText &BERT, XLNet, RoBERTa & Title, abstract\\
\cellcolor{blue!20}\cite{kamateri2022automated}  & FastText, Glove, Word2Vec  &CNN, LSTM, GRU & TADC\\
 \cellcolor{green!20}\cite{ghauri2023classification}& Vision Transformer  &MLP & Image\\
 \cellcolor{blue!20}\cite{kamateri2023ensemble}     & FastText &Bi-LSTM, Bi-GRU, LSTM   & Metadata\\
 \cellcolor{blue!20}\cite{bekamiri2024}     & SBERT &KNN & Claim, title, abstract\\

\bottomrule

\end{tabular}
}
\vspace{-1mm}
\label{tab:patent_classification}
\end{table*}

\subsubsection{Traditional Neural Networks}
The commonality among these methods is that they follow a two-step approach: generate initial features and then use a classifier for the final classification. One of the initial studies \cite{grawe2017automated} implements a single-layer LSTM to classify patents at the IPC subgroup level where the initial features are obtained by the Word2Vec method. Similarly, \cite{shalaby2018lstm} use LSTM for IPC subclass level classification. For the initial document representation, the method uses fixed hierarchy vectors that utilize distinct models for various segments of the document. \cite{risch2018learning} and \cite{risch2019domain} focus on training fastText word embeddings on a corpus of 5 million patent documents, then use Bi-GRU for classification. Similarly, \cite{sofean2021deep} applies text mining techniques to extract key sections from patents, train Word2Vec, and then use multiple parallel LSTMs for the classification task. These collectively show the usefulness of neural networks in patent classification.

\subsubsection{Ensemble Models}
The models in this category are used to ensemble different word embeddings and deep learning models. \cite{benites2018classifying} use SVM as a baseline method and experiment with various datasets, the number of features, and semi-supervised learning approaches. Meanwhile, \cite{kamateri2023ensemble} and \cite{kamateri2022automated} both investigate ensemble models incorporating  Bi-LSTM, Bi-GRU, LSTM, and GRU. More specifically,  \cite{kamateri2022automated} conduct experiments with different word embedding techniques, whereas \cite{kamateri2023ensemble} focus on applying various partitioning techniques to enhance the performance of the proposed framework. While the above methods heavily focus on texts,  \cite{ghauri2023classification} classify patent images into distinct types of visualizations, such as graphs, block circuits, flowcharts, and technical drawings, along with various perspectives, including side, top, left, and perspective views. The approach utilizes the CLIP model with Multi-layer Perceptron (MLP) and various CNN models.

\begin{table*}[ht]
\label{class}
\caption{Works on patent retrieval. The papers are white, blue, and gray based on the data type of text, image, and both, respectively. The dataset details are provided in Appendix \ref{app:datasets}.}
\centering
\resizebox{0.9\textwidth}{!}{%
\begin{tabular}{cccc}
\toprule
\textbf{Work}  & \textbf{Method} &\textbf{Training}&\textbf{Datasets}\\ 
\hline 
\cellcolor{blue!20}\cite{Kravets2017/12}     & CNN& supervised& Freepatent, Findpatent \\
 \cite{kang2020patent}        & BERT& pre-trained & WIPS\\
\cite{chen2020deep}  & BiLSTM-CRF, BiGRU-HAN& supervised& USPTO \\
\cellcolor{blue!20} \cite{jiang2021deriving}     & DUAL-VGG& supervised&-\\

\cite{setchi2021artificial}    & SVM, Naive Bayes, Random Forest, MLP& supervised& -\\ 
\cellcolor{gray!20}\cite{pustu2021multimodal}         & RoBERTa, CLIP& pre-trained& EPO\\
\cite{siddharth2022enhancing}   & Sentence-BERT, TransE& pre-trained, unsupervised &USPTO\\
\cellcolor{blue!20}\cite{Kucer_2022_WACV}   &  (ImageNet, Sketchy) ResNet50& supervised, finetuned&DeepPatent\\
\cellcolor{blue!20}\cite{higuchi2023patent}  & Deep Metric Learning& self-supervised&DeepPatent\\
\cellcolor{blue!20}\cite{higuchi2023patent2}  & InfoNCE and ArcFace& self-supervised &DeepPatent\\
\cellcolor{blue!20}\cite{lo2024large}  & BLIP-2, GPT-4V & pre-trained, supervised &DeepPatent2\\
\bottomrule

\end{tabular}
}
\label{tab:patent_retrieval}
\end{table*}

\subsubsection{Pre-trained Language Models (PLMs)}
The first study \cite{2lee2020patent} which involves PLMs, fine-tune the BERT model on the USPTO-2M dataset and introducing a new dataset, USPTO-3M at the subclass level to aid in future research. Concurrently, \cite{haghighian2022patentnet} also fine-tune BERT, along with XLNet \cite{xlnet}, and RoBERTa on the USPTO-2M dataset. They establish XLNet as the new state-of-the-art in classification performance, achieving the highest precision, recall, and f1 measure. \cite{althammer2021linguistically} implement domain adaptive pre-trained 
Linguistically Informed Masking and shows that SciBERT-based representations perform better than BERT-based representations in patent classification. SciBERT is pre-trained on scientific literature which helps the method to understand the technical language of patents. \cite{bekamiri2024} use Sentence BERT that takes into account entire sentences instead of word by word. On USPTO data, their method gives the highest recall and f1 score.
\subsubsection{Discussion and Suggestion }The evaluation measures for patent classification are accuracy, precision, recall, and the f1 score on the CPC or IPC. The earlier works on patent classification are mostly focused on simpler neural networks \cite{risch2018learning,risch2019domain}. Applying models such as LSTM can capture the sequence and context in the text, which is suitable for the patent domain since the context is critical. However, these are comparatively simple models that might be limited to capturing complex technical structures in patent documentation. This limitation is evident in the evaluation metrics; for instance, the highest accuracy at the subclass level is only 0.74 (Table \ref{tab:patent_classification_results} in Appendix). More advanced techniques, including PLMs, have become popular over time. PLMs could be powerful because of their pretraining step on a massive amount of data. Patent text is different from the usual text in scientific articles (e.g., research papers). Thus, fine-tuning PLMs on patent datasets might be able to address some of these concerns by providing context-aware representations for the patent domain. From Table \ref{tab:patent_classification_results}, the early works have a low precision of 0.53 on USPTO data \cite{risch2018learning}. PLMs--- such as BERT and RoBERTa---have significantly improved the performance to 0.82 \cite{haghighian2022patentnet}. The language models used for classification tasks in the patent domain are generally simpler compared to advanced LLMs such as GPT and LLaMA. There is a significant gap between recent practices in the patent domain and the existing advanced AI models. However, direct performance comparisons across methods are limited by  differences in dataset subsets, class hierarchies, and evaluation metrics used in different studies.

\begin{table*}[ht]
\label{class}
\caption{Summary of the methods on patent quality: ``Many" includes Linear regression, Ridge regression, Random Forest, XGBoost, CNN, and LSTM. ``APR" stands for the measures of accuracy, precision, and recall. IncoPat is a global patent database. We denote Attribute Network Embedding, Attention-based Convolutional Neural Network, European Telecommunications Standards Institute, Derwent Innovation by ANE, ACNN, ETSI, and DI, respectively.  
}
\centering
\resizebox{\textwidth}{!}{
\begin{tabular}{ccccc}
\toprule
\textbf{Papers} & \textbf{Indicators} & \textbf{Methods} &\textbf{Evaluation Metrics } &\textbf{Datasets}\\ 
\hline 
\cite{lin2018patent} & Citations, meta features & ANE, ACNN & RMSE & USPTO, OECD \\
\cite{trappey2019patent}  &  Principal component analysis (PCA)   & DNN & Accuracy & ETSI and DI \\
\cite{hsu2020deep}      & Investor
reaction, citations& Many & MAE & Patentsview    \\
\cite{chung2020early}& Abstract, claims, predefined &CNN, Bi-LSTM & Precision, recall & USPTO \\

\cite{aristodemou2021identifying}& 12 patent indices & ANN & APR, F1, FNR, MAE & USPTO, OECD \\
\cite{erdogan2022predicting} & 9 patent indices & MLP & Accuracy, Kappa, MAE & USPTO  \\ 
\cite{li2022deep}  & Maintenance period & BiLSTM-ATT-CRF  & APR, F1 & IncoPat \\
\cite{krant2023text}  & Patent text & MSABERT & MSE& USPTO, OECD \\
\bottomrule

\end{tabular}
}
\vspace{-1mm}
\label{tab:quality}
\end{table*}

\subsection{Patent Retrieval}
We organize the relevant studies below based on the types of methods. Table \ref{tab:patent_retrieval} provides an overview of studies for patent retrieval. We present the results by these methods in Table \ref{tab:patent_retrieval_results} (Appendix \ref{app:eval}).

\subsubsection{Traditional Machine Learning} Initial studies have used traditional machine learning methods for patent retrieval. \cite{setchi2021artificial} describe five technical requirements to investigate the feasibility of AI for the task. These requirements include query expansion and identification of semantically similar documents. The study uses SVMs, Naive Bayesian learning, decision tree induction, and RF, along with word embeddings, to solve the prior art retrieval problem. Prior art usually implies the references which may be used to determine the novelty of a patent application. Patent data is searched through multiple resources and returns results based on the query and the database and these results need to be merged to create the final result. \cite{stamatis2023machine} employ techniques such as random forest, Support Vector Regression, and Decision Trees to merge the search findings effectively.

\subsubsection{Traditional Neural Networks}
The methods based on neural networks have been popular in recent years for patent retrieval. \cite{Kravets2017/12}, \cite{jiang2021deriving}, and \cite{Kucer_2022_WACV} implement CNN, DUAL-VGG, and ResNet, respectively, to retrieve patent images based on a query image. \cite{chen2020deep} aim to solve entity identification and semantic relation extraction by BiLSTM-CRF \cite{huang2015bidirectional} and BiGRU-HAN \cite{han2019opennre}, respectively.

\subsubsection{PLMs \& Multimodal Models (MMs)}
PLMs are useful in many text-related tasks and patent retrieval is not an exception. \cite{kang2020patent} use the BERT language model which includes the combinations of title, abstract, and claim. \cite{siddharth2022enhancing} incorporate Sentence-BERT \cite{sbert} for text embeddings as well as use the TransE method for the citation and inventor knowledge graph embeddings. They identify that the mean cosine similarity among the vector representations of the patents is effective in linking multiple existing patents to a target patent. Multimodal techniques have also been used in information retrieval \cite{pustu2021multimodal}. Here, the visual features are extracted using vision transformers, while textual features are from sentence transformers. \cite{pustu2021multimodal} utilize CLIP for image embedding alongside RoBERTa for capturing textual features, and thus, enhances the search process by incorporating both visual and textual data. \cite{lo2024large} use distribution-aware contrastive loss to improve understanding of class and category information which achieves robust representations even for tail classes. For captioning, they employ open-source BLIP-2 and GPT-4V, a frozen text encoder from CLIP for text feature, and various visual encoder backbones, including ViT variants, ResNet50, EfficientNetB-0, and SwinV2-B. 
Among other techniques, \cite{higuchi2023patent}, \cite{higuchi2023patent2} employ a deep metric learning framework with cross-entropy methods such as InfoNCE \cite{oord2018representation} and ArcFace \cite{deng2019arcface}.

\subsubsection{Discussion and Suggestion }Patent retrieval process involves several subtasks, such as defining technical requirements and merging search outcomes from various databases. The early methods often use traditional techniques like SVM, Naive Bayes, Decision trees, etc. While the image retrieval methods apply a variety of CNNs to effectively handle and analyze the visual data, the text retrieval methods have shifted towards PLMs for advanced linguistic analysis. Traditional machine learning techniques are limited to capturing the complexity of both patent image and text. Although CNNs are popular for image retrieval tasks, the question remains in their effectiveness for patent image retrieval, as patent images are non-traditional and technical. On the other hand, combining Vision Transformer alongside RoBERTa, Sentence-BERT, TransE shows another approach that might be more suitable for handling the multimodal (e.g., text, images) aspect of patents. \cite{pustu2021multimodal} demonstrate that the image and text-based transformer models achieve the highest mean average precision in patent retrieval tasks. Table \ref{tab:multimodal-comparison} provides a comparative overview of multimodal approaches, fusion strategies and dataset sizes.

\begin{table*}[ht]
\vspace{-2mm}
\scriptsize
\label{tab:llm-trend}
\caption{Example of the works that used PLMs and LLMs to solve patent tasks. This shows the growing trend of incorporating large-scale language models to improve patent processing and analysis.}
\centering
\resizebox{0.78\textwidth}{!}{%
\begin{tabular}{cccccccc}
\toprule
\textbf{Work} &  \textbf{Model} &\textbf{Task} &\textbf{Year}\\ 
\hline 
\cite{2lee2020patent} & BERT& Classification & 2020\\
\cite{kang2020patent}& BERT& Retrieval & 2020\\
\cite{lee2020patent} & GPT-2 & Generation & 2020\\
\cite{lee2020patent2} & GPT-2 & Generation& 2020\\
\midrule
\cite{althammer2021linguistically} & SciBERT & Classification & 2021\\
\cite{pustu2021multimodal}& RoBERTa & Retrieval& 2021\\
\midrule
\cite{haghighian2022patentnet} & BERT, RoBERTa & Classification& 2022\\
\cite{siddharth2022enhancing}   & SBERT & Retrieval & 2022\\
\cite{christofidellis2022pgt} & GPT-2 & Generation& 2022\\
\midrule
\cite{krant2023text}   & MSABERT & Quality Analysis & 2023\\
\midrule
\cite{bekamiri2024} & Sentence-BERT & Classification  & 2024\\
\cite{lo2024large}   & BLIP, GPT-4 & Retrieval & 2024\\
\cite{wang2024patentformer}& GPT-J, T5 & Generation& 2024\\
\cite{instructpatentgpt}& GPT-J & Generation& 2024\\
\cite{jiang2024can}& Llama-3, Mistral, and PatentGPT-J& Generation& 2024\\
\cite{bai2024patentgpt}& Llama-2 and Mixtral& Generation& 2024\\
\cite{ren2024patentgpt}& Qwen2& Generation& 2024\\
\cite{wang2024autopatent}& Qwen2, LLAMA3, GPT-4o, Mistral& Generation& 2024\\
\bottomrule
\end{tabular}
}
\vspace{-1mm}
\label{tab:llm-trend}
\end{table*}

\subsection{Patent Quality Analysis}

We organize the methods for patent quality analysis below and provide a summary in Table \ref{tab:quality}.

\subsubsection{Traditional Neural Networks}
\cite{erdogan2022predicting} apply an MLP-based approach for quality analysis, utilizing nine indices such as claim counts, forward citations, backward citations, the patent family size to measure the value of a patent, etc. \cite{li2022deep} classify patents based on their maintenance period in four categories. This study implements a Bi-LSTM along with the attention mechanism and Conditional Random Field (CRF) to predict the quality of a patent. \cite{trappey2019patent} use Deep Neural Networks with 11 quality indicators. \cite{hsu2020deep}  predict forward citation and investor reaction to patent announcements implementing CNN-LSTM neural networks and various ML models. \cite{chung2020early}, \cite{lin2018patent} and \cite{aristodemou2021identifying} apply a variety of neural networks such as CNN, Bi-LSTM, Attention-based CNN (ACNN), deep and wide Artificial Neural Networks (ANN), respectively.
\subsubsection{Pre-trained Language Models (PLMs)}
\cite{krant2023text} proposes to use MSABERT to assess patent value based entirely on the textual data and use the OECD \cite{OECD} quality indicators for evaluation. Building upon BERT, MSABERT handles the multi-section structure and longer texts of patent documents. The OECD index includes composite indicators and generality with other predominant indices. 

\subsubsection{Discussion and Suggestion}
While numerous measures are used in assessing the quality of a patent, the absence of universally accepted \quotes{gold standard} poses a challenge. Among several used indices, only forward citations are directly associated with the value---both monetary and quality---of a patent. Even though applying different deep learning models has some success, the question of building a method to handle technical information, metadata, and images together remains open. While MSABERT on the entire dataset will be computationally costly, building upon it might be useful for quality evaluation.

\vspace{-2mm}

\subsection{Patent Generation}
\label{sec:generation}

The generative models are becoming increasingly popular in many domains. The recent developments in LLMs have also led to novel methods for generating patents, thus reducing significant human effort.
Sec. \ref{sub:gen} presents the studies with LLMs and PLMs for generating patent texts, and Sec. \ref{sub:pgen} focuses on the pretrained and  advanced methods used for patent-specific data. 
Table \ref{tab:llm-trend} shows the trend of using PLMs and LLMs to solve different patent tasks, and most patent-related tasks are shifting towards leveraging LLMs. Table \ref{tab:generation} (see Appendix) shows the summary of patent generation. We also discuss the broader impact in App.~\ref{app:impact}.

\subsubsection{Patent Text Generation with LLMs}
\label{sub:gen}
\cite{lee2020patent} implement GPT-2 \cite{gpt2} models to generate the independent claims in patents. The researchers fine-tune the model on 555,890 patent claims of the granted utility patents in 2013 from USPTO. Providing a few words, the method generates the first independent claim of the patent. However, the study is limited to providing quantitative metrics to evaluate the quality of the generated patent claims. In a separate study, \cite{lee2020patent2} focuses on personalized claim generation by fine-tuning a pre-trained GPT-2 model with inventor-centric data to demonstrate greater relevance. The measure of personalization in the generated claims has been assessed using a BERT model. \cite{christofidellis2022pgt} introduce the Patent Generative Transformer (PGT) that supports three tasks: part-of-patent generation, text infilling, and coherence evaluation. They train GPT-2 on a dataset of 11.6 million patents. PGT shows strong zero-shot capabilities for generating abstracts with high semantic similarities from keywords. Patentformer \cite{wang2024patentformer}  generates detailed patent specifications by fine-tuning T5 and GPT-J language models on a dataset that includes claims, drawings, and descriptions. It focuses on two tasks: Claim-to-Specification, which creates specification text from a single claim, and Claim+Drawing-to-Specification, which integrates claims, drawings, and descriptions to produce richer specifications. \cite{jiang2024can} generate claims by incorporating descriptions instead of abstracts. It also demonstrates an interesting observation that the general-purpose models---such as Llama-3, GPT-4, and Mistral---outperform models specifically trained on patent data (e.g., PatentGPT-J). The authors also conclude that fine-tuning enhances clarity, but revisions are still necessary for legal robustness.

\subsubsection{Patent-Specific LLMs}
\label{sub:pgen}
\cite{instructpatentgpt} finetunes a pretrained model PatentGPT-J-6B using reinforcement learning from human feedback (RLHF) to align patent claim generation with drafting goals. The authors design a custom reward function where claim length up to a defined length and inclusion of limiting terms are rewarded. These limiting terms improve the chance of patent approval. However, further improvements in text quality and broader datasets are needed to meet legal and practical patent standards. \cite{bai2024patentgpt} build a cost-effective LLMs for the intellectual property (IP) domain to handle domain-specific expertise and long-text processing. They finetune open-source models like LLaMA2 and Mixtral with over 240 billion multilingual IP-focused tokens, nearly half from patent data. The approach incorporates pretraining, fine-tuning, and reinforcement learning to align model outputs with human preferences. Similarly, \cite{ren2024patentgpt} introduce a specialized LLM based on Qwen2-1.5b for automated patent drafting. The approach integrates domain-specific knowledge using knowledge graphs, supervised fine-tuning, and RLHF. A multi-agent framework for drafting patents using LLMs is introduced by \cite{wang2024autopatent}. They employ agents for planning, writing, and reviewing to generate comprehensive patents from inventor drafts. 

\subsubsection{Discussion and Suggestion } 
The use of PLMs and LLMs for automating patent generation has grown rapidly. However, a critical challenge remains in evaluating the quality of generated patents. Interestingly, general purpose models have outperformed domain specific models \cite{jiang2024can} in this task. This outcome may reflect the stronger generalization and linguistic capacity of larger open models. The existing studies focus only on pretraining LLMs on patent-specific data to better capture the domain’s technical language and structure without rigorous evaluation techniques. As a result, human intervention becomes essential to ensure accuracy, legal validity, and compliance with patent standards. Additionally, most approaches for patent generation focus exclusively on the text and overlook the multimodal nature of patents. This is particularly important for design patents, which consist of images predominantly. \\


\section{Future Directions}
\label{sec:future}
Many researchers have leveraged NLP and Multimodal AI for patent analysis, yet significant research opportunities remain going forward. We believe a foundation model (e.g., LLMs, MLMs) tailored for patent data will enhance understanding and performance across diverse tasks.\\

\vspace{-1mm}
\textbf{Multimodal Learning on Patents.} The availability of multiple modalities (e.g., text, images) in patent documents offers a comprehensive understanding of the related patent tasks. One of the challenges is that the patent images are often more complex and use advanced domain related concepts compared to the natural (or RGB) images. Recent advances in multimodal learning would allow for more reliable and accurate patent analysis. Intuitively, drawings or sketches provide geometrical information about individual patents. In general, multimodal learning can be used to \textit{align representations} derived from text descriptions with those derived from technical images. 

\textbf{Generative AI for Patents. }In patent generation, LLMs can suffer from hallucination, where they generate incorrect information. They might produce repetitive and monotonous texts that will lack creativity. Further, to mitigate the risk of patent infringement, LLMs need up-to-date patent data. Thus, the generation process requires human oversight and feedback to ensure accuracy and relevance and cannot be fully automated yet.  On the other hand, the assessment of the text generated by the generative models is also challenging. As patents include jargons and many domain specific words, evaluating generated patent text in terms of only natural language will not be sufficient. Thus, the important question remains---\textit{how to construct domain-specific evaluation measures for the synthetic or the generated text from LLMs?}

Some prior works adopt automatic metrics such as BLEU or ROUGE to evaluate patent generation. However, we acknowledge that these are insufficient for assessing the factual accuracy or legal correctness of the generated patent text. Promising directions to address this include:
(i) Using retrieval-based evaluation to check consistency with prior art. By comparing generated patent content against existing patents, models can better ensure novelty and reduce the risk of infringement;
(ii) Applying retrieval-augmented generation (RAG) to improve grounding and factual accuracy. RAG enables the model to retrieve and reference relevant patent documents or technical literature at generation time, making the draft more contextually aligned and reliable;
(iii) Applying reinforcement learning from human feedback (RLHF) to reduce hallucinations and increase legal robustness. In this setting, patent experts (e.g., attorneys or reviewers) can rate or correct generated claims and descriptions based on novelty, clarity, and consistency with prior art. This structured feedback can guide models to avoid generating unsupported technical features or inaccurate functionalities.

\textbf{Patent Assessment.} To asses patent's novelty, one of the major tasks is to retrieve similar patents to determine whether the patent is significantly different from existing patents. One of the important task in this case is to generate search queries. This often needs alternate search terms, related words, and synonyms which require domain knowledge. The quality and structure of queries directly impact the relevance of the search results. The current methods are yet to automate this entire process. Thus, it brings challenges to obtain adequate similar patents and correctly assess patent's innovativeness and novelty. On the other hand, the generic quality analysis are based on well-known measures \cite{aristodemou2021identifying,erdogan2022predicting}. As an example, patent citation has been considered as a proxy for patent valuation \cite{nandi2024experimental,hsu2020deep}. Specifically, these works involve prediction of patent value dependent on citation count from the text. Nonetheless, it remains unclear which of these indices are associated with the actual value of the patent (e.g., generated revenue).

\textbf{Building a Knowledge Graph.} Patents are represented as nodes connected by edges such as citations in a citation network \cite{liu2022early}. This structured representation allows for detailed citation analysis which is considered a crucial metric in understanding a patent's value. One interesting future direction would be to build a knowledge graph using other important information such as meta-data, semantic similarity of patents, etc. This may lead to a more organized landscape of patents. This knowledge graph can help with prior art searches, the identification of related patents, and identify valuable patents (e.g., patents with high citations) \cite{siddharth2022enhancing}.

\textbf{Cross-jurisdictional Retrieval.} An important and largely unexplored direction is cross-jurisdictional patent retrieval, such as between USPTO and EPO corpora. These tasks introduce additional challenges arising from differences in legal terminology, language, classification codes, and document formatting across jurisdictions. We highlight these as promising future direction to enhance the generalizability and robustness of patent retrieval systems.
We believe that integrating these techniques into future generative frameworks will enhance their reliability, reduce hallucinated content, and align model outputs more closely with legal and technical standards in patent systems.

\section{Conclusions} 

In this survey, we have provided a comprehensive overview of various patent analysis tasks. We have presented a novel schema with a detailed organization of the research papers, analyzing the corresponding methodologies, their advantages, limitations, and how they are applied to different patent-related tasks. Our survey also focuses on the recent advancements of PLMs and LLMs as well as their usefulness in the patent domain. We have offered several insights into some potential future directions. This survey aims to be a useful guide for researchers, practitioners, and patent offices all over the world in the multidisciplinary field of NLP, Multimodal AI, and patent systems.

\clearpage
\section{Limitations}
\vspace{-2mm}
The life cycle of a patent---the time from its submission to acceptance---is lengthy as it undergoes significant scrutiny and multiple iterations of revisions. The advancements in Machine Learning (e.g., LLMs) can make this process faster and thus, can essentially accelerate technological innovation. For instance, while reviewing, recent tools can help retrieve relevant documents more efficiently and accurately than a human reviewer who often requires enough experience. Our work is a survey of the existing methods for such tasks in patents. Though the survey itself does not have limitations as such, we discuss the limitations of modern AI techniques in general for patent tasks.

There are a few limitations of using AI in patent analysis. First, the LLMs methods may lack the nuanced understanding that human experts possess. Second, evaluation scores in classification and retrieval indicate lower accuracy (see Tables \ref{tab:patent_classification_results} \& \ref{tab:patent_retrieval_results}) and thus, they still need human intervention to obtain relevant literature---which is important while reviewing---to prevent the patent infringement issues. Therefore, the entire process cannot be fully automated, and it is important to have human experts in the loop. This requirement also applies to generative models for patent drafting (Sec.~\ref{sec:generation}) which needs human guidance for accuracy. Additionally, there are ethical concerns regarding the potential displacement of human workers by AI tools. 

\section{Ethics Statement}
In this work, we have surveyed AI methods for patent tasks. We do not foresee any ethical issues from our study.
\clearpage
\bibliography{custom}

\clearpage
\appendix
\clearpage
\section{Appendix}
\subsection{Overview of the tasks}
\label{app:overview}
\begin{figure*}[ht]
\centering
\includegraphics[width=0.95\textwidth]{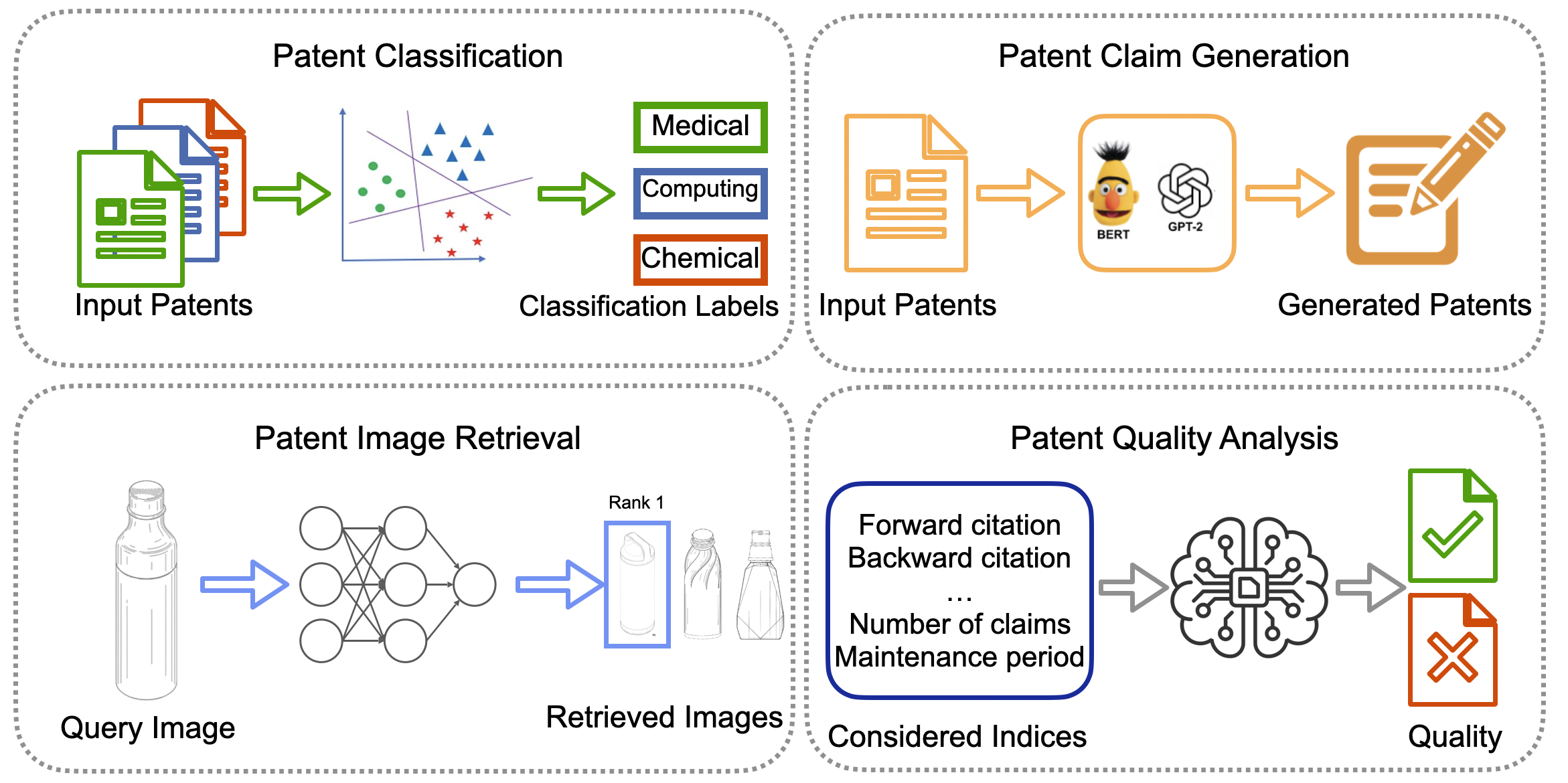}

\caption {\label{fig:overview}%
The overview of four major tasks of patent analysis. The patent retrieval task includes obtaining relevant patents (text and images). Please refer to the detailed descriptions of these tasks in Section \ref{sec:background}.}
\end{figure*}

Overview of the major patent tasks: patent classification, patent retrieval, patent generation, and patent quality analysis is shown in Figure \ref{fig:overview}. Popular AI methods in the literature covered by this survey are listed in Table \ref{tab:summary_methods}.
\subsection{Search and inclusion criteria. }
\label{app:search}
We have conducted our literature search using Google Scholar and Semantic Scholar, focusing on various categories of patent-related tasks. To align with the recent trends, we have limited our search to publications from 2017 to 2024. Our search criteria included various keywords such as `patent', `AI in patent', `patent classification', `patent tasks', `patent retrieval', `patent generation', `patent quality analysis', and `patent dataset'. This combination of search terms has yielded hundreds of patent-related research papers. We have excluded more than half of these papers after reviewing their titles and abstracts, as they have not met our criteria (e.g., they did not fall under any of the relevant categories). After thorough scrutiny and reorganization, we have included 50 papers for the survey. \\

\subsection{Background: Formulation of Patent Tasks}
\label{app:formulation}
We provide the problem formulations of the popular patent tasks as follows. 

\subsubsection{Patent Classification}
Given patents as $(x_i, y_{i})_{i=1}^{N}$, where $x_i$ denotes the features of the  $i$-th patent, $C$ denotes the set of classes, $C=\{1,2,\dots,k\}$, and $y_i=\{y_{i1},y_{i2},\dots,y_{iK}\}$ is a binary multi-label vector, where $y_{ik} \in\{ 0,1\}$ is an indicator of whether class $k$ is the correct classification for the example patent $i$. Since a single patent can belong to more than one class in \( C \), the goal is to predict \( y_{i} \).\\
Table \ref{tab:classification_example} shows an example of CPC classification.

\begin{table*}[ht]
\label{class}
\caption{Summary of the works on patent generation. Here, "comprehensive" denotes patent claims, specification drafting, classification, translation, etc. IP data includes research papers, litigation records, web, news, etc.   
}
\centering
\resizebox{\textwidth}{!}{
\begin{tabular}{ccccc}
\toprule
\textbf{Papers} & \textbf{Model} &\textbf{Parts} &\textbf{Data}\\ 
\hline 
\cite{lee2020patent} & GPT-2 & Independent Claims & USPTO  \\
\cite{lee2020patent2} & GPT-2 & Personalized Claims & USPTO  \\
\cite{christofidellis2022pgt} & GPT-2  & title, abstract, claim& --  \\
\cite{wang2024patentformer} & T5, GPT-J (Patentformer)&Claim-to-Specification, Claim+Drawing-to-Specification  & USPTO  \\
\cite{jiang2024can} & Llama-3, Mistral, and PatentGPT-J & Claims & HUPD \\
\cite{instructpatentgpt} & PatentGPT-J & Claim & USPTO, PatentsView  \\
\cite{bai2024patentgpt} & LLaMA2 and Mixtral & Comprehensive & Both patent and IP data  \\
\cite{ren2024patentgpt} &Qwen2 &Comprehensive & USPTO  \\
\cite{wang2024autopatent}&Qwen2, LLAMA3, GPT-4o, Mistral& Comprehensive & HUPD\\
\bottomrule
\end{tabular}
}
\label{tab:generation}
\end{table*}

\begin{table*}[ht]
\small
\caption{Popular AI methods in the literature. We use the acronyms frequently in our survey.}
\centering
\begin{tabular}{ccc}
\toprule
\textbf{Acronym} & \textbf{Full Name} & \textbf{Paper} \\ \hline 
LSTM  &  Long short-term memory    &   \cite{LSTM}  \\ 
CNN &  Convolutional Neural Networks     &  \cite{CNN}   \\ 

Bi-LSTM &  Bidirectional Long Short-Term Memory    &  \cite{bi-lstm}   \\
Word2Vec&   --   &   \cite{mikolov2013efficient}  \\ 
GRU   &   Gated Recurrent Units   &  \cite{GRU}   \\ 
Bi-GRU &  Bidirectional Gated Recurrent Units    &  \cite{GRU} \\ 

DUAL-VGG & Dual Visual Geometry Group &\cite{vgg} \\
FastText & -- & \cite{fastText}   \\
BERT& Bidirectional Encoder Representations from Transformers     &    \cite{BERT} \\ 
RoBERTa& Robustly Optimized BERT Pre-training Approach &\cite{roberta}\\ 
SciBERT & Scientific BERT & \cite{scibert} \\ 
\bottomrule
\end{tabular}
\label{tab:summary_methods}
\end{table*}

\begin{table*}[ht]
\small
\caption{An example of Cooperative Patent Classification (CPC) Scheme for the section A and its hierarchical categorization.}
\centering
\begin{tabular}{lll}
\toprule
\textbf{Level} & \textbf{Code} & \textbf{Category} \\ \hline 
Section  &  A & Human Necessities\\ 
Class  &  A61 & Medical or Veterinary Science: Hygiene \\ 
Sub-class  &  A61B & Diagnosis: Surgery: Identification\\ 
Group  &  A61B5 & Measuring for diagnostic purposes; Identification of persons \\ 
Sub-group  &  A61B5/0006 & ECG or EEG signals \\ 
\bottomrule
\end{tabular}
\label{tab:classification_example}
\end{table*}

\begin{table*}[ht]
\caption{Existing results on the patent classification task. Hierarchy levels for classification include Section, Class, Subclass, Group, and Subgroup. The tuple (Result 1, Reuslt 2) denotes the results using (Data 1, Data 2) for the papers that report the measures using multiple datasets separately. The WIPO-alpha is a dataset for automated patent classification systems, and ALTA2018 is a dataset from Language Technology Programming Competition.}
\centering
\small
\resizebox{\textwidth}{!}{%
\begin{tabular}{cccccccc}\toprule
\textbf{Papers} & \textbf{Hierarchy Level} & \textbf{Accuracy} &\textbf{Precision}&\textbf{Recall} &\textbf{F1} &\textbf{Top-3} &\textbf{Data}\\ 
\hline 
\cite{grawe2017automated} & Subgroup & 0.63 & 0.63& 0.66 & 0.62 & -- &USPTO\\ 
\cite{shalaby2018lstm}&  Subclass & -- &--& -- & 0.61 & 0.79: F1 & - \\
\cite{shalaby2018lstm}&  Class & -- &--& -- & 0.72 & 0.89: F1& - \\
\cite{risch2018learning} &  Subclass  & -- &(0.49, 0.53)& -- & -- & (0.72,0.75): Precision & WIPO-alpha, USPTO\\
\cite{benites2018classifying}   & Class& -- &--& -- & 0.78 & -- & ALTA2018, WIPO \\
\cite{risch2019domain}     &  Subclass & -- &(0.49, 0.53)& -- & -- & (0.72,0.75): Precision & WIPO-alpha, USPTO \\
\cite{2lee2020patent}& Subclass & -- &0.81& 0.55 & 0.65 & 0.44: F1 & USPTO \\
\cite{althammer2021linguistically}& Subclass &  0.59 & 0.58 & 0.59 & 0.581 & -- & USPTO\\
\cite{sofean2021deep} & Subclass & 0.74 & 0.92 & 0.63 & 0.75 & -- & EPO, WIPO \\
\cite{haghighian2022patentnet} & Subclass & -- & (0.82, 0.82) & (0.55, 0.67) & (0.63, 0.72)& --&USPTO, CLEF-IP 2011\\
\cite{kamateri2022automated}  & Subclass & 0.64 & --& -- & -- & -- &CLEF-IP 2011 \\
\cite{ghauri2023classification}&  Image type &0.85 & --& -- & -- & -- & CLEF-IP 2011, USPTO\\
\cite{kamateri2023ensemble}     & Subclass& 0.68 & --& -- & -- & 0.89: accuracy &CLEFIP-0.54M\\
\cite{bekamiri2024}     & Subclass& -- & 0.67& 0.71 & 0.66 & --&USPTO\\

\bottomrule

\end{tabular}
}
\label{tab:patent_classification_results}
\end{table*}

\begin{table*}[ht]
\label{class}
\caption{Results of the papers for the Patent Retrieval task. Here, mAP denotes mean average precision. Freepatent and Findpatent are patent data websites, where Findpatent includes patents registered in Russia. WIPS is a patent information search system.}
\centering
\resizebox{\textwidth}{!}{%
\begin{tabular}{cccccccc}
\toprule
\textbf{Work} &  \textbf{Data type} &\textbf{Data} &\textbf{Accuracy (\%)} &\textbf{Precision}&\textbf{Recall} &\textbf{F1} &\textbf{mAP}\\ 
\hline 
\cite{Kravets2017/12} & image & Freepatent, Findpatent &30&--&--&--&--\\
 \cite{kang2020patent} & text & WIPS&-- & 71.74 & 94.29 & 81.48&--  \\
\cite{chen2020deep}  & text& USPTO& -- &92.4 & 91.9 & 92.2 &-- \\
\cite{pustu2021multimodal}& image+text& EPO & --& --& --& --&0.715 \\
\cite{siddharth2022enhancing}& text&USPTO & 70.2 & 65.9 & 81.2 & 72.6 & -- \\
\cite{Kucer_2022_WACV}   & image& DeepPatent & 70.1 &--&--&--& 37.9 \\
\cite{higuchi2023patent}  & image& DeepPatent &--&--&--&--& 0.85\\
\cite{higuchi2023patent2}  & image& DeepPatent& --&--&--&--& 0.622\\
\cite{lo2024large}  & image& DeepPatent2& --&--&--&--& 0.69\\
\bottomrule

\end{tabular}
}
\label{tab:patent_retrieval_results}
\end{table*}

\begin{table*}[h!]
\small
\caption{Overview of Patent Datasets: size, format, data type and intended tasks}
\centering\
\begin{tabular}{ccccc}
\toprule
\textbf{Dataset} & \textbf{Size} & \textbf{Format}  & \textbf{Data type} & Task\\ 
\hline 
USPTO-2M \cite{li2018deeppatent}  & 2M & JSON &   text  & Classification   \\ 
BIGPATENT \cite{sharma2019bigpatent}   &  1.3M &   JSON &text & Summarization  \\
USPTO-3M \cite{2lee2020patent}  & 3M & SQL statement &  text & Classification \\
 PatentMatch \cite{risch2021patentmatch}    & 6.3M &  JSON &  text & Retrieval \\
DeepPatent  \cite{Kucer_2022_WACV}  & 350K &  XML \& PNG &  text \& image & Retrieval\\
DeepPatent2 \cite{Ajayi_2023}    &  2M & JSON \& PNG &  text \& image & Retrieval \\

HUPD \cite{harvard}  & 4.5M &  JSON &  text & Multi-purpose \\
IMPACT \cite{shomee2024impact}& 3.61M & CSV \& TIFF & text \& image & Multi-purpose \\
\bottomrule

\end{tabular}
\label{table1}
\end{table*}

\subsubsection{Patent Retrieval}
 
Given a query patent as $q$ and a set of patents $X=\{x_i,\ldots,x_n\}$, where $x_q$ and $x_i$ are the features of the query and the patent $i$ in the set $X$. The goal is to compute a similarity score (e.g. cosine) $s(x_q, x_i)$  and return a set of patents $R(q)=\{x_j,\ldots,x_k\}$ based on top-k high similarities.

\subsubsection{Patent Generation}
 Given the patent $x_i$, where $x_i$ are the features constructed from the instruction, title, abstract, or any other part of the patent of the example patent $i$, the output $y_i$ can be another part of the patent (e.g., abstract, the first claim). The generation function $G$ can be denoted as $y_i$=$G(x_i;\theta)$, 
where $\theta$ is the parameter of the generation model $G$. The goal is to generate $y_i$ by learning $\theta$, or inferring from a pre-trained model with learned $\theta$.

Table \ref{tab:generation} shows the summary of the models and datasets used to generate parts of the patent text.

\subsection{Evaluation results}
\label{app:eval}
We discuss all the studies and related methods in Section \ref{sec:methods}. We present the evaluation metrics and the results in Table \ref{tab:patent_classification_results} and \ref{tab:patent_retrieval_results}.
\subsection{NLP and AI-based Methods for Other Relevant Patent Tasks}
\label{app:other}
There are other interesting studies in the patent domain. Recent work focuses on patent infringement, such as \cite{chi2022establish} develop a model with different deep learning methods, such as CNN and LSTM, to predict the possibility of a patent application being granted and classify the reason for a failed application. Another work \cite{choi2022deep} applied a transformer and a Graph Neural Network (GNN) on patent classification for patent landscaping. \cite{zaini2022identifying} present an unsupervised method to identify the correlations between patent classification codes and search keywords using PCA and k-means. These studies provide advanced deep learning methods to avoid the risks in patent application. Moreover, there are various studies on generating new ideas and evaluating novelty, such as identifying the inventive process of novel patent using BERT~\cite{giordano2023unveiling}, and an explainable AI (XAI) model for novelty analysis via \cite{jang2023explainable}. \cite{zou2023event} propose a new task to predict the trends of patents for the companies, and also provide a solution for the task by training an event-based GNN. These studies bring new insights and directions for patent ideas and developments.\\
\textbf{Applications in Businesses.}
The use of LLMs among businesses for patent related processes has significantly risen over time. The usage of the machine learning methods for these patents is growing at an impressive average annual rate of 28\%\footnote{\url{https://ip.com/blog/can-ai-invent-independently-how-ai-is-changing-the-patent-industry/}}. Businesses are increasingly applying AI to enhance various aspects of the patent process, from drafting and classification to search and analysis. Some of the prominent examples include \cite{qatent}, \cite{davinciai}, and \cite{questel}. \cite{qatent} leverages the latest NLP techniques to facilitate  patent drafting for patent practitioners. It focuses on automating routine tasks---typing, automating renumbering of claims, and antecedence checking. It recommends various word and sentence alternatives during the claim drafting process, such as synonyms, broader or more specific terms, and other linguistic variations. Despite recent discussions around AI-generated inventions, Qatent maintains a human-centric approach which ensures all outputs are driven and controlled by human drafters.
\cite{davinciai} is an advanced tool for drafting patents that uses generative AI to streamline the process. It supports a variety of document formats and lets users alter the AI's writing style to suit their needs.
\cite{questel} offers AI powered patent classification, comprehensive patent search capabilities, efficient exploration of new markets, and opportunities such as management of patent fees and renewals.\\

\subsection{Patent Dataset and Repositories}
\label{app:datasets}
Patent data are publicly available for bulk download from several sources in various formats such as XML, TSV, TIFF, and PDF. Examples include the USPTO,  PatentsView\footnote{\url{https://patentsview.org/}}, EPO\footnote{\url{https://www.epo.org/}}, and WIPO\footnote{\url{https://www.wipo.int/classifications/ipc/en/ITsupport/}}. Freepatent and Findpatent are patent data websites, where Findpatent includes patents registered in Russia. Beyond these resources, several patent datasets are available for benchmarking purposes. The datasets are detailed in Table \ref{table1}. 

\begin{table*}[ht]
\small
\caption{Comparative overview of multimodal methods with fusion strategies and dataset sizes.}
\centering
\label{tab:multimodal-comparison}
\begin{tabular}{p{2cm} p{2.cm} p{3cm} p{1cm} p{3.2cm} p{1.8cm}}
\toprule
\textbf{Paper} & \textbf{Model} & \textbf{Data Size} & \textbf{Task} & \textbf{Fusion Strategy} & \textbf{Modalities} \\
\midrule
Pustu-Iren et al. (2021) & CLIP + RoBERTa & 30,379 patent images & Retrieval & Late fusion (separate text/image encoders) & Text + Image \\
Lo et al. (2024) & BLIP-2 + ViT + GPT-4 & 822,792 images (train) & Retrieval & Contrastive alignment (dual encoders + InfoNCE loss) & Text + Image \\
\bottomrule
\end{tabular}
\end{table*}

\subsection{Broader Impacts}
\label{app:impact}
The life-cycle of a patent---the time from its submission to acceptance---is lengthy as it undergoes significant scrutiny and multiple iterations of revisions. The advancements in LLMs can make this process faster and thus, can essentially accelerate technological innovation. For instance, while reviewing, recent tools can help retrieve relevant documents more efficiently and accurately than a human reviewer who often requires enough experience.\\
Some of the major benefits are as follows: (1) Speed: The inclusion of LLMs and Multimodal AI in patent analysis tasks will speed up the review process. For example, \cite{ghauri2023classification} use a vision transformer that classifies images much more efficiently than previous works, and \cite{bekamiri2024} achieve higher recall in classification tasks. Since patent classification is a time-consuming task for a human expert, incorporating these advancements into the review process will make the process faster.
(2) Novelty: Another important task is retrieving similar patents which is essential to assess the novelty of a patent. \cite{higuchi2023patent} show a satisfactory mAP in retrieving similar images, which can play a key role in patent infringement. (3)Innovation: \cite{lee2020patent,lee2020patent2} explore generating new patents, which is an important component to foster new innovation. Their work provides inspiration for further development in the field including creation of new and innovative patents.

\end{document}